\documentclass[conference,a4paper]{IEEEtran}
\usepackage[utf8]{inputenc}
\usepackage[T1]{fontenc}
\usepackage{url}
\usepackage{ifthen}
\usepackage{cite}
\usepackage[cmex10]{amsmath}
\usepackage{graphicx}
\usepackage{amssymb,bm}
\usepackage{bbm}
\usepackage{amsmath}
\usepackage{amsfonts}
\usepackage{cite}
\usepackage{enumitem}
\DeclareMathOperator{\E}{\mathbbmss{E}}

\allowdisplaybreaks
\usepackage{amsthm}
\newtheorem{theorem}{Theorem}

\newtheorem{lemma}{Lemma}
\newtheorem{remark}{Remark}

\usepackage{chngcntr}
\counterwithout{theorem}{section}
\counterwithout{definition}{section}
\counterwithout{lemma}{section}
\counterwithout{remark}{section}
\counterwithout{assumption}{section}
\counterwithout{proposition}{section}
\counterwithout{corollary}{section}
\IEEEoverridecommandlockouts
\begin{document}
\bstctlcite{IEEEexample:BSTcontrol}
\title{On Multi-Cell Uplink-Downlink Duality with Treating Inter-Cell Interference as Noise}
 \author{\IEEEauthorblockN{Hamdi~Joudeh\IEEEauthorrefmark{1}, Xinping~Yi\IEEEauthorrefmark{2} and Bruno~Clerckx\IEEEauthorrefmark{1}}
   \IEEEauthorblockA{\IEEEauthorrefmark{1}Department of Electrical and Electronic Engineering, Imperial College London\\
                     \IEEEauthorrefmark{2}Department of Electrical Engineering and Electronics, University of Liverpool\\
                     Email: \IEEEauthorrefmark{1}\{hamdi.joudeh10, b.clerckx\}@imperial.ac.uk, \IEEEauthorrefmark{2}xinping.yi@liverpool.ac.uk}
\thanks{H. Joudeh and B. Clerckx were supported in part by the U.K. Engineering and Physical Sciences Research Council (EPSRC) under grant EP/N015312/1.}}

\maketitle
\begin{abstract}
We consider the information-theoretic optimality of treating inter-cell interference as noise in downlink cellular networks modeled as Gaussian interfering broadcast channels.
Establishing a new uplink-downlink duality, we cast the problem in Gaussian interfering broadcast channels to that
in Gaussian interfering multiple access channels, and characterize an achievable GDoF region under power control and
treating inter-cell interference as (Gaussian) noise.
%
We then identify conditions under which this achievable GDoF region is optimal.
\end{abstract}
\section{Introduction}
\label{sec:introduction}
The age-old robust interference management strategy of power control and treating interference as noise (TIN)
in wireless networks has recently been given renewed vitality, attracting increasing attention of late
\cite{Geng2015,Geng2015a,Geng2016,Sun2016,Yi2016,Gherekhloo2016,Gherekhloo2017,Yi2018,Joudeh2018a}.
This revived interest is largely due to the findings of Geng \emph{et al.} \cite{Geng2015}, who showed that power control and TIN are
sufficient to achieve the entire generalized degrees-of-freedom (GDoF) region, and capacity region to within a constant gap,
of the $K$-user Gaussian interference channel (IC) in a broad regime of parameters, described in terms of channel strength levels.
The approach and results of Geng \emph{et al.} were generalized and extended in a number of directions reported in \cite{Geng2015a,Geng2016,Sun2016,Yi2016,Gherekhloo2016,Gherekhloo2017,Yi2018,Joudeh2018a}.

The TIN framework of \cite{Geng2015} was recently extended to multi-cell networks in uplink scenarios \cite{Joudeh2018a},
modeled as the Gaussian interfering multiple access channel (IMAC).
TIN is defined in \cite{Joudeh2018a} for such setting as the employment of ``\emph{a MAC-type, capacity-achieving strategy, with
Gaussian codebooks and successive decoding....in each cell while treating all inter-cell interference as noise....complemented with power control to manage inter-cell interference}''.
Under this TIN scheme, the achievable GDoF region (with no time-sharing) was explicitly characterized as a finite union of polyhedra,
and broad regimes, with respect to channel strength levels, in which this region is a polyhedron and optimal were identified \cite{Joudeh2018a}.
A natural question then arises as to whether this multi-cell TIN framework, constructed for uplink cellular
networks, is also valid for their downlink counterparts.
In this paper, we make progress towards answering this question.

We consider the downlink counterpart of the uplink setting in \cite{Joudeh2018a},
modeled by the Gaussian interfering broadcast channel (IBC) \cite{Suh2011}, comprising $K$ mutually interfering Gaussian BCs.
We compose the downlink counterpart of the TIN scheme in \cite{Joudeh2018a}, i.e.
each cell employs a power-controlled, degraded BC-type, capacity-achieving strategy, with superposition coding and successive decoding,
while treating inter-cell interference as (Gaussian) noise.
We establish a new uplink-downlink duality between IBCs and IMACs in the sense that
the corresponding TIN-achievable GDoF regions are unchanged if the roles of transmitters
and receivers are switched.
This duality, in conjunction with the results in \cite{Joudeh2018a}, leads to an explicit characterization of the
TIN-achievable GDoF region for the IBC.
We further identify conditions under which this IBC achievable GDoF region is also optimal.

\emph{Notation:}
For positive integers $z_{1}$ and $z_{2}$ where $z_{1} \leq z_{2}$, the sets $\{1,2,\ldots,z_{1}\}$ and $\{z_{1},z_{1}+1,\ldots,z_{2}\}$ are denoted by
$\langle z_{1} \rangle$ and $\langle z_{1}:z_{2}\rangle$, respectively.
For any $a \in \mathbb{R}$, $(a)^{+} = \max\{0,a\}$.
Bold lowercase symbols  denote tuples, e.g. $\mathbf{a} = (a_{1},\ldots,a_{Z})$.
For $\mathcal{A} = \{\mathbf{a}_{1},\ldots,\mathbf{a}_{K}\}$, $\Sigma(\mathcal{A})$
is the set of all cyclicly ordered sequences of all subsets of $\mathcal{A}$ (see \cite[Sec. 1.3]{Joudeh2018a}).
\section{System Model}
Consider a $K$-cell cellular network in which each cell $k$, where $k \in \langle K \rangle$, comprises a base station
denoted by BS-$k$ and two user equipments, each denoted by UE-$(l_{k},k)$, where $l_{k} \in \langle 2 \rangle$.
The set of tuples corresponding to all UEs in the networks is given by
$\mathcal{K} \triangleq \left\{(l_{k},k) : l_{k} \in \langle 2 \rangle, k \in \langle K \rangle  \right\}$.
For ease of exposition, we limit our attention to the $2$-user-per-cell case.
Nevertheless, the results can be generally extended to scenarios with an arbitrary number of users in each cell.
\subsection{Interfering Broadcast Channel}
\label{subsec:IBC_system_model}
\begin{figure}
\centering
\includegraphics[width = 0.45\textwidth,trim={5.5cm 5.1cm 5.5cm 5.2cm},clip]{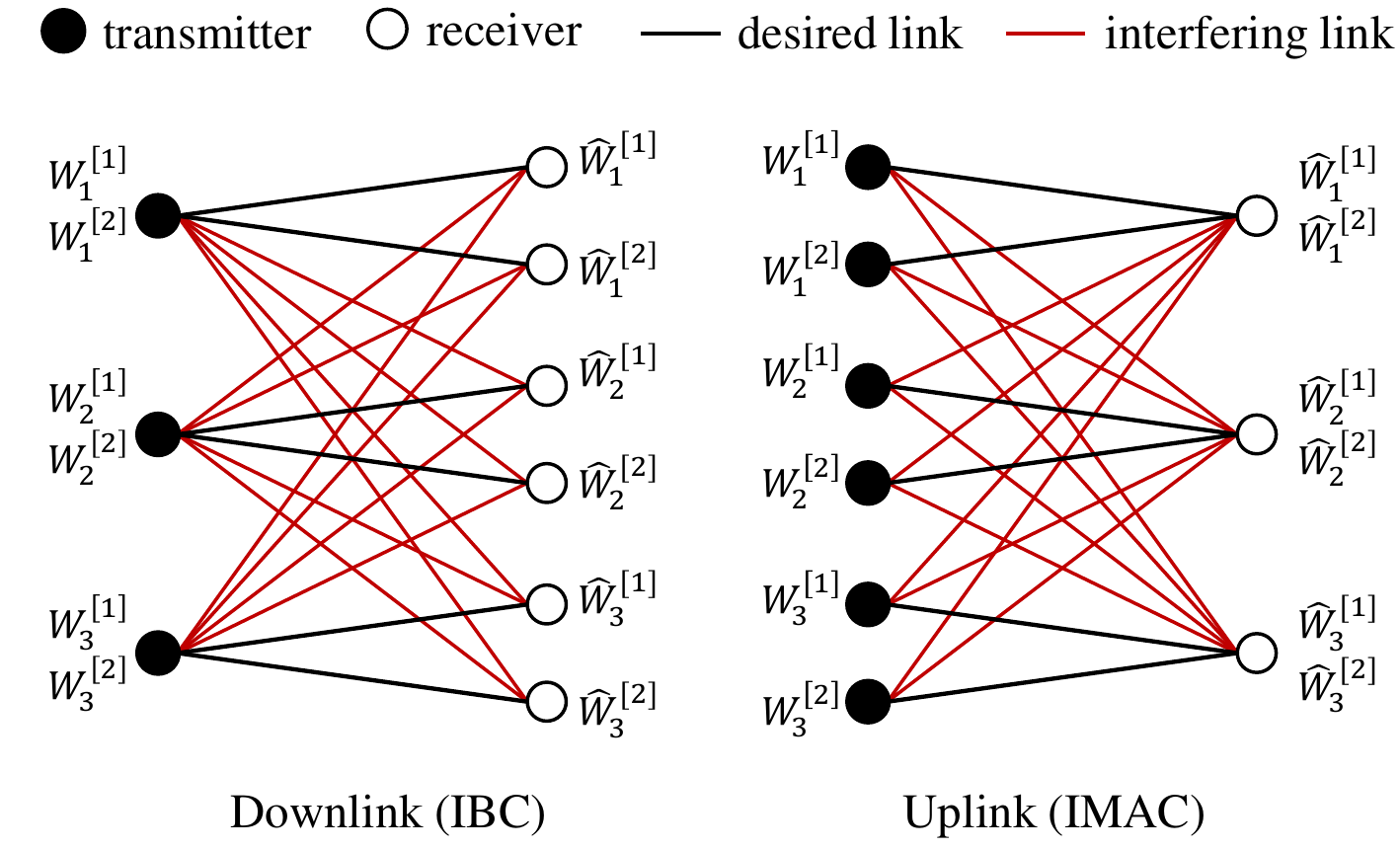}
\caption{A 3-cell interfering broadcast channel (donwlink) and its dual interfering multiple access channel (uplink).}
\label{fig:IBC_IMAC_example}
\vspace{-2mm}
\end{figure}

When operating in the downlink mode, the above network is modeled by a Gaussian IBC, e.g. Fig. \ref{fig:IBC_IMAC_example} (left).
Adopting a GDoF-friendly model (see \cite{Geng2015}), the input-output relationship at the $t$-th use of the channel, where $t \in \mathbb{N}$, is described as
\begin{align}
\nonumber
Y_{k}^{[l_{k}]}(t)
& = \sum_{i = 1}^{K} h_{ki}^{[l_{k}]} X_{i}(t)
+ Z_{k}^{[l_{k}]}(t)\\
\label{eq:IBC_system model 2}
& = \sum_{i = 1}^{K} \sqrt{ P^{\alpha_{ki}^{[l_{k}]}} } e^{j \theta_{ki}^{[l_{k}]}} X_{i}(t)
+ Z_{k}^{[l_{k}]}(t),
\end{align}
where $Y_{k}^{[l_{k}]}(t)$ is the signal received by UE-$(l_{k},k)$, $h_{ki}^{[l_{k}]}$ is the channel coefficient from BS-$i$ to UE-$(l_{k},k)$,
$X_{i}(t)$ is the transmitted symbol of BS-$i$ and $Z_{k}^{[l_{k}]}(t) \sim \mathcal{N}_{\mathbb{C}}(0,1)$
is the additive white Gaussian noise (AWGN) at UE-$(l_{k},k)$.
All symbols are complex and each BS-$i$ is subject to the average power constraint
$\frac{1}{n}\sum_{t=1}^{n}\E \big[|X_{i}(t)|^{2}\big] \leq 1$, where $n$ is the duration of the communication.
$\sqrt{ P^{\alpha_{ki}^{[l_{k}]}} }$ and $\theta_{ki}^{[l_{k}]}$ are the magnitude and phase of the channel coefficient $h_{ki}^{[l_{k}]}$,
where $P > 0$ is a nominal power value and $\alpha_{ki}^{[l_{k}]} \geq 0$ is the corresponding channel strength level\footnote{As in \cite{Geng2015}, avoiding negative strength levels has no impact on the results.}.
Without loss of generality, we assume that the following order holds
\begin{equation}
\label{eq:strength_order}
\alpha_{kk}^{[1]}  \leq  \alpha_{kk}^{[2]}, \ \forall k \in \langle K \rangle.
\end{equation}
Each transmitter in the IBC, e.g. BS-$k$ for some $k \in \langle K \rangle$,
has the independent messages $W_{k}^{[1]}$ and $W_{k}^{[2]}$
intended to UE-$(1,k)$ and UE-$(2,k)$, respectively.
Codes, error probabilities, achievable rates
$\mathbf{R}= \big(R_{1}^{[1]},R_{1}^{[2]},\ldots,R_{K}^{[1]},R_{K}^{[2]}\big)$,
and the capacity region $\mathcal{C}^{\mathrm{IBC}}$
are all defined in the standard Shannon theoretic sense.
A GDoF tuple is denoted by $\mathbf{d} = \big(d_{1}^{[1]},d_{1}^{[2]},\ldots,d_{K}^{[1]},d_{K}^{[2]}\big)$
and the GDoF region is denoted by $\mathcal{D}^{\mathrm{IBC}}$, where both are defined in the standard fashion.
\subsection{Dual Interfering Multiple Access Channel}
The dual IMAC is obtained by reversing the roles of the transmitters and receivers in the IBC,
e.g. Fig. \ref{fig:IBC_IMAC_example} (right).
Building upon the GDoF-friendly model in \eqref{eq:IBC_system model 2}, the input-output
relationship for the IMAC is given by
\begin{align}
\label{eq:IMAC_system model 2}
\bar{Y}_{i}(t)  = \sum_{k = 1}^{K}  \Big[ h_{ki}^{[1]} \bar{X}_{k}^{[1]}(t) + h_{ki}^{[2]} \bar{X}_{k}^{[2]}(t) \Big]
+ \bar{Z}_{i}(t),
\end{align}
where $\bar{Y}_{i}(t)$ and $\bar{Z}_{i}(t) \sim \mathcal{N}_{\mathbb{C}}(0,1)$ are the received signal and the AWGN at BS-$i$
respectively, and $\bar{X}_{k}^{[l_{k}]}(t)$ is the transmitted symbol of UE-$(l_{k},k)$.
Each UE-$(l_{k},k)$ is subject to the average power constraint
$\frac{1}{n}\sum_{t=1}^{n}\E \big[|\bar{X}_{k}^{[l_{k}]}(t)|^{2}\big] \leq 1$.
For cell $k$, $k \in \langle K \rangle$,  UE-$(1,k)$ and UE-$(2,k)$ have the independent messages $W_{k}^{[1]}$ and $W_{k}^{[2]}$, respectively,
intended to BS-$k$.
We denote the capacity region and the GDoF region of the above IMAC by $\mathcal{C}^{\mathrm{IMAC}}$ and $\mathcal{D}^{\mathrm{IMAC}}$, respectively.
\begin{remark}
When defining dual uplink channels, it is common to impose a sum transmit power constraint on the UEs so that the total transmit power does not exceed the transmit power of the BS in the downlink channel.
We do not impose this on the dual IMAC here, which exhibits a transmit
power gain of $K$. This gain is inconsequential for the GDoF results.
\end{remark}
\section{Treating Inter-cell Interference as Noise and Uplink-Downlink Duality}
We consider TIN in the cellular sense \cite{Joudeh2018a}, where each cell employs
an adequately modified single-cell capacity-achieving strategy while treating all
inter-cell interference as noise.
\subsection{TIN in the IBC}
\label{subsec:TIN_IBC}
Each cell $k$ of the IBC employs superposition coding
and successive decoding according to an order $\pi_{k}$, which is a permutation on the set $\{1,2\}$.
In particular, the transmitted signal of BS-$k$  is composed as
\begin{equation}
X_{k}(t) =  X_{k}^{[1]}(t) + X_{k}^{[2]}(t),
\end{equation}
where each message $W_{k}^{[l_{k}]}$, for $l_{k} \in \langle 2 \rangle$, is independently encoded into the codeword $X_{k}^{[l_{k}]n} \triangleq  X_{k}^{[l_{k}]}(1),\ldots,X_{k}^{[l_{k}]}(n) $, drawn from a Gaussian codebook
with average power $\frac{1}{n}\sum_{t=1}^{n}\E \big[ | X_{k}^{[l_{k}]}(t) |^{2} \big] = q_{k}^{[l_{k}]}$.
Note that the powers $q_{k}^{[1]}$ and $q_{k}^{[2]} $ satisfy $q_{k}^{[1]} + q_{k}^{[2]} \leq 1$.
On the other end,  UE-$\big(\pi_{k}(1) \big)$ decodes its own signal $X_{k}^{[\pi_{k}(1)]n}$ while treating all other signals (i.e. both intra-cell and inter-cell interference) as noise.
UE-$\big(\pi_{k}(2) \big)$, however, starts by decoding and cancelling $X_{k}^{[\pi_{k}(1)]n}$ before decoding its own signal
$X_{k}^{[\pi_{k}(2)]n}$, while treating inter-cell interference as noise.

Using the above scheme, the effective signal-to-interference-plus-noise ratio (SINR)
for decoding the signal $X_{k}^{[\pi_{k}(1)]n}$ is denoted by $\mathrm{SINR}_{k}^{[\pi_{k}(1)]}$ (given by a minimum of two SINRs),
while the SINR for $X_{k}^{[\pi_{k}(2)]n}$ is denoted by $\mathrm{SINR}_{k}^{[\pi_{k}(2)]}$
(SINR expressions are omitted for brevity).
For fixed decoding order and power allocation, the message $W_{k}^{[\pi_{k}(l_{k})]}$ is hence reliably communicated to
UE-$\big(\pi_{k}(l_{k}),k\big)$ at any rate satisfying
\begin{equation}
\label{eq:IBC_rate per user}
0  \leq  R_{k}^{[\pi_{k}(l_{k})]} \leq   \log
\Big(  1+  \mathrm{SINR}_{k}^{[\pi_{k}(l_{k})]}   \Big).
\end{equation}

For GDoF purposes, we may assume that $q_{k}^{[l_{k}]}$
scales with $P$ as $O(P^{r_{k}^{[l_{k}]}})$, where $r_{k}^{[l_{k}]} \leq 0$
is the corresponding transmit power exponent.
It follows that the achievable GDoF tuple in cell $k$ is given by all $d_{k}^{[\pi_{k}(1)]}$
and $d_{k}^{[\pi_{k}(2)]}$ that satisfy \eqref{eq:IBC_GDoF per user} (top of next page).
\begin{figure*}[!t]
\vspace{-2mm}
\normalsize
\begin{subequations}
\label{eq:IBC_GDoF per user}
\begin{align}
&0 \leq d_{k}^{[\pi_{k}(1)]}  \leq  \max \biggl\{  0, \min_{m_{k} \in \{1,2\} }    \Bigl\{  \alpha_{kk}^{[\pi_{k}(m_{k})]}   +   r_{k}^{[\pi_{k}(1)]}
 -   \Bigl(  \max \bigl\{ \alpha_{kk}^{[\pi_{k}(m_{k})]}  +  r_{k}^{[\pi_{k}(2)]} ,\max_{(l_{j},j):j \neq k} \{ \alpha_{kj}^{[\pi_{k}(m_{k})]}
 +   r_{j}^{[l_{j}]} \} \bigr\}  \Bigr)^{+}  \Bigr\}   \biggr\} \\
& 0 \leq d_{k}^{[\pi_{k}(2)]}  \leq \max \biggl\{ 0, \alpha_{kk}^{[\pi_{k}(2)]} + r_{k}^{[\pi_{k}(2)]}
 -  \Bigl( \max_{(l_{j},j): j \neq k} \{   \alpha_{kj}^{[\pi_{k}(2)]} + r_{j}^{[l_{j}]}  \} \Bigr)^{+} \biggr\}.
\end{align}
\end{subequations}
\hrulefill
\vspace{-2mm}
\end{figure*}
A power allocation tuple is given by
$\mathbf{r}  =  \big(r_{1}^{[1]},r_{1}^{[2]},\ldots,r_{K}^{[1]},r_{K}^{[2]} \big) \leq \mathbf{0}$
and a network decoding order tuple is given by $\bm{\pi} \triangleq \left( \pi_{1},\ldots,  \pi_{K} \right) \in \Pi$,
where $\Pi$ is the set comprising all possible network decoding orders.
For fixed $(\bm{\pi},\mathbf{r})$, we use $\mathcal{P}_{\bm{\pi}}^{\mathrm{IBC}\star}(\mathbf{r})$ to denote
the set of all GDoF tuples   $\mathbf{d}$ with components satisfying \eqref{eq:IBC_GDoF per user}.
The \emph{TIN-achievable GDoF region} with fixed $\bm{\pi} \in \Pi$ is obtained by taking the union over all
feasible power allocations, i.e.
$\mathcal{P}_{\bm{\pi}}^{\mathrm{IBC}\star} \triangleq  \bigcup_{\mathbf{r} \leq \mathbf{0}}
\mathcal{P}_{\bm{\pi}}^{\mathrm{IBC}\star}(\mathbf{r})$.
The \emph{general TIN-achievable GDoF region} for the IBC, denoted by $\mathcal{P}^{\mathrm{IBC}\star}$, is obtained by further considering all possible decoding orders in $\Pi$ and is defined as
\begin{equation}
\label{eq:IBC_TIN_GDoF_region}
\mathcal{P}^{\mathrm{IBC}\star} \triangleq \bigcup_{\bm{\pi} \in \Pi} \mathcal{P}_{\bm{\pi}}^{\mathrm{IBC}\star}.
\end{equation}
As time-sharing is not allowed, each GDoF tuple $\mathbf{d} \in \mathcal{P}^{\mathrm{IBC}\star}$ is achieved through a strategy identified by a decoding order and a power allocation tuple, i.e. $(\bm{\pi},\mathbf{r})$.
\begin{figure*}[!t]
\vspace{-3mm}
\normalsize
\begin{subequations}
\label{eq:IMAC_GDoF per user}
\begin{align}
&0 \leq \bar{d}_{k}^{[\pi_{k}(1)]}  \leq
\max \biggl\{  0,   \alpha_{kk}^{[\pi_{k}(1)]} +  \bar{r}_{k}^{[\pi_{k}(1)]}
  -  \Bigl(\max_{(l_{j},j): j \neq k} \{ \alpha_{jk}^{[l_{j}]} + \bar{r}_{j}^{[l_{j}]}   \}    \Bigr)^{+} \biggr\}
\\
\label{eq:IMAC_GDoF per user_2}
&0 \leq \bar{d}_{k}^{[\pi_{k}(2)]}  \leq
\max \biggl\{   0,   \alpha_{kk}^{[\pi_{k}(2)]}  +   \bar{r}_{k}^{[\pi_{k}(2)]}
  -  \Bigl(\max \Bigl\{ \alpha_{kk}^{[\pi_{k}(1)]} + \bar{r}_{k}^{[\pi_{k}(1)]} ,
  \max_{(l_{j},j): j \neq k} \{ \alpha_{jk}^{[l_{j}]} + \bar{r}_{j}^{[l_{j}]}   \}   \Bigr\} \Bigr)^{+}  \biggr\}.
\end{align}
\end{subequations}
\hrulefill
\vspace{-3mm}
\end{figure*}
\subsection{TIN in the Dual IMAC}
For the dual IMAC, we adopt the TIN scheme given in \cite{Joudeh2018a}.
We highlight the aspects most relevant to this paper.
Readers are referred to \cite{Joudeh2018a} for a more detailed exposition.

Each UE-$(l_{k},k)$ uses an independent Gaussian codebook with power $P^{r_{k}^{[l_{k}]}}$, where the transmit power exponent
$\bar{r}_{k}^{[l_{k}]}$ and the power allocation tuple $\bar{\mathbf{r}} \leq \mathbf{0}$ are define similarly to their counterparts in the IBC.
Each BS-$k$ successively decodes its in-cell signals according to the order
$\pi_{k}$,
such that $\bar{X}_{k}^{[\pi_{k}(2)]}$ is decoded and cancelled before decoding $\bar{X}_{k}^{[\pi_{k}(1)]}$.
As for the IBC, the network decoding order tuple is given by $\bm{\pi} \in \Pi$.
For a decoding order $\bm{\pi}$ and a power allocation $\bar{\mathbf{r}}$,
UE-$\big(\pi_{k}(l_{k}),k\big)$ achieves any GDoF $\bar{d}_{k}^{[\pi_{k}(l_{k})]}$ satisfying \eqref{eq:IMAC_GDoF per user} (top of this page).

As in the IBC, while fixing $(\bm{\pi},\bar{\mathbf{r}})$,
we achieve $\mathcal{P}_{\bm{\pi}}^{\mathrm{IMAC}\star}(\bar{\mathbf{r}})$ given by all
all GDoF tuples $\bar{\mathbf{d}}$ with components satisfying \eqref{eq:IMAC_GDoF per user}.
The TIN-achievable GDoF region, for fixed $\bm{\pi}$, is given by
$\bigcup_{ \bar{\mathbf{r}} \leq \mathbf{0}} \mathcal{P}_{\bm{\pi}}^{\mathrm{IMAC}\star}(\bar{\mathbf{r}})$,
while the general TIN-achievable GDoF region for the dual IMAC is given by
\begin{equation}
\label{eq:IMAC_TIN_GDoF_region}
\mathcal{P}^{\mathrm{IMAC}\star} \triangleq \bigcup_{\bm{\pi} \in \Pi} \mathcal{P}_{\bm{\pi}}^{\mathrm{IMAC}\star}.
\end{equation}
\begin{remark}
For any given cell $k$ and permutation $\pi_{k}$, the uplink decoding order
is the reverse of the counterpart downlink decoding order. This reverse relationship is commonly
exhibited in uplink-downlink dualities (see \cite[Ch. 10.3.4]{Tse2005}).
\end{remark}
\subsection{Uplink-Downlink Duality under TIN}
Here we present the first result of this paper.
\begin{theorem}
\label{theorem:TIN_UL_DL_Duality}
The IBC and IMAC general TIN-achievable GDoF regions $\mathcal{P}^{\mathrm{IBC}\star}$ and $\mathcal{P}^{\mathrm{IMAC}\star}$
are identical.
\end{theorem}
To prove Theorem \ref{theorem:TIN_UL_DL_Duality}, we first consider an arbitrary IBC GDoF tuple $\mathbf{d} \in \mathcal{P}^{\mathrm{IBC}\star}$.
From the earlier parts of this section, we know that there must exist a decoding order $\bm{\pi} \in \Pi$
and a feasible power allocation tuple $\mathbf{r} \leq \mathbf{0}$
such that  $\mathbf{d} \in \mathcal{P}_{\bm{\pi}}^{\mathrm{IBC}\star}(\mathbf{r})$.
We show that for the same $\bm{\pi}$, there exists $\bar{\mathbf{r}} \leq \mathbf{0}$
such that $\mathbf{d} \in \mathcal{P}_{\bm{\pi}}^{\mathrm{IMAC}\star}(\bar{\mathbf{r}})$ also holds.
This proves that $ \mathcal{P}^{\mathrm{IBC}\star} \subseteq  \mathcal{P}^{\mathrm{IMAC}\star} $ in general, as the selected GDoF tuple $\mathbf{d}$ is arbitrary.
We then apply a similar argument in the opposite direction and prove that
$\mathcal{P}^{\mathrm{IMAC}\star} \subseteq \mathcal{P}^{\mathrm{IBC}\star}$.
Details of this proof are presented in Appendix \ref{appebdix:proof_of_duality}.
It is worthwhile noting that a similar uplink-downlink duality of achievable GDoF tuples and power allocations
was shown for the regular $K$-user IC in \cite{Geng2018}.

From the duality in Theorem \ref{theorem:TIN_UL_DL_Duality} and the characterization of $\mathcal{P}^{\mathrm{IMAC}\star}$ in
\cite[Th. 2]{Joudeh2018a}, we obtain a
characterization of $\mathcal{P}^{\mathrm{IBC}\star}$  as a finite union of polyhedra.
Moreover, from \cite[Th. 3]{Joudeh2018a}, we obtain conditions under which $\mathcal{P}^{\mathrm{IBC}\star}$
is a polyhedron, i.e. one of the polyhedra in the union includes all others.
\section{On TIN-Optimality for the IBC}
In the following theorem, we obtain \emph{TIN-optimality conditions} under which the TIN scheme described in Section \ref{subsec:TIN_IBC}
achieves the entire GDoF region of the IBC.
\begin{theorem}
\label{theorem:TIN_optimality}
For the IBC described in Section \ref{subsec:IBC_system_model}, if the following conditions are satisfied
\begin{align}
\label{eq:TIN_condition_1}
\alpha_{ii}^{[2]}  & \geq  \alpha_{ii}^{[1]} + \max_{j:j\neq i} \left\{ \alpha_{ij}^{[2]} \right\}, \
\forall i \in \langle K \rangle \\
\label{eq:TIN_condition_2}
\alpha_{ii}^{[l_{i}]} & \geq \max_{j:j \neq i} \left\{\alpha_{ij}^{[l_{i}]} \right\} +
\max_{(l_{k},k):k \neq i} \left\{\alpha_{ki}^{[l_{k}]} \right\}, \ \forall(l_{i},i) \in \mathcal{K},
\end{align}
then the optimal GDoF region is given by  $\mathcal{D}^{\mathrm{IBC}} = \mathcal{P}^{\mathrm{IBC}\star}$,
which is characterized by all GDoF tuples that satisfy
\begin{align}
\label{eq:IBC_TIN_region_1}
d_{i}^{[l_{i}]}   & \geq 0, \ \forall (l_{i},i) \in \mathcal{K} \\
\label{eq:IBC_TIN_region_2}
\sum_{s_{i} \in \langle l_{i} \rangle} d_{i}^{[s_{i}]} & \leq \alpha_{ii}^{[l_{i}]}, \ \forall l_{i} \in \{1,2\}, i \in \langle K \rangle \\
\nonumber
\sum_{j \in \langle m \rangle }  \! \sum_{s_{i_{j}} \! \in \! \langle l_{i_{j}} \rangle}
\! \! \! d_{i_{j}}^{[s_{i_{j}}]} &  \! \leq \! \! \! \!
\sum_{j \in \langle m \rangle } \alpha_{i_{j}i_{j}}^{[l_{i_{j}}]} \! - \alpha_{i_{j}i_{j-1}}^{[l_{i_{j}}]}, \ \forall l_{i_{j}} \in \{1,2\}, \\
\label{eq:IBC_TIN_region_3}
 (i_{1},\ldots,i_{m})   & \in \Sigma(\langle K \rangle), m \in \langle 2:K \rangle.
\end{align}
\end{theorem}
It is readily seen that the IBC TIN-optimality conditions identified in Theorem \ref{theorem:TIN_optimality}
imply (i.e. stricter than) the counterpart IMAC TIN-optimality conditions identified in \cite[Th. 4]{Joudeh2018a}.
From this observation, it follows that $\mathcal{P}^{\mathrm{IBC}\star}$ is characterized by
\eqref{eq:IBC_TIN_region_1}--\eqref{eq:IBC_TIN_region_3} under such conditions.
The rest of the section is hence dedicated to proving the converse part of Theorem \ref{theorem:TIN_optimality}.

We start with an auxiliary result that plays a key role in the proof.
This result essentially shows that under condition \eqref{eq:TIN_condition_1}, UE-$(2,k)$
is \emph{more capable} than UE-$(1,k)$ in a GDoF sense.
\subsection{Auxiliary Lemma}
Consider the independent input sequences $X_{1}^{n}$ and $X_{2}^{n}$ with average power constraints
$\frac{1}{n}\sum_{t=1}^{n}\E \big[|X_{i}(t)|^{2}\big] \leq 1$, $i \in \{1,2\}$.
Let $Y_{a}^{n}$ and $Y_{a}^{n}$ be noisy outputs given by
\begin{align}
Y_{a}(t) & = a_{1}X_{1}(t) + a_{2}X_{2}(t) + Z_{a}(t)  \\
Y_{b}(t) & = b_{1}X_{1}(t) + b_{2}X_{2}(t) + Z_{b}(t)
\end{align}
where $a_{1},a_{2},b_{1},b_{2} \in \mathbb{C}$ are constants and $Z_{a}(t),Z_{b}(t) \sim\mathcal{N}_{\mathbb{C}}(0,1)$ are AWGN terms.
Finally, let $W$ be a discrete random variable such that $W \rightarrow X_{1}^{n} \rightarrow Y_{a}^{n}$
and $W \rightarrow X_{1}^{n} \rightarrow Y_{b}^{n}$ are Markov chains.
The following result holds.
\begin{lemma}
\label{lemma:I_diff}
Given that the two conditions $\frac{ |b_{1}|^{2} }{ |b_{2}|^{2}  } \geq |a_{1}|^{2}$
and $ |b_{2}|^{2}  \geq |a_{2}|^{2} \geq 1$
hold, then
\begin{equation}
\label{eq:lemma_I_diff}
I(X_{1}^{n} ; Y_{a}^{n} | W )  \leq I(X_{1}^{n} ; Y_{b}^{n} | W ) + n.
\end{equation}
\end{lemma}
The proof of the above result is relegated to Appendix \ref{appebdix:proof_lemma_I_diff}.
\subsection{Proof of Theorem \ref{theorem:TIN_optimality}}
For each cell $i$, the inequalities in \eqref{eq:IBC_TIN_region_2} follow from the capacity region of the degraded Gaussian BC \cite{ElGamal2011}. Therefore, we focus on the cyclic bounds in \eqref{eq:IBC_TIN_region_3}.
Cells and users participating in a given cyclic bound are identified by the sequences
$(i_{1},\ldots,i_{m}) \in \Sigma\big(\langle K \rangle\big)$ and $(l_{i_{1}},\ldots,l_{i_{m}}) \in \{1,2\}^{m}$.
Next, we go through the two following steps:
\begin{itemize}
\item Eliminate all non-participating UEs, all non-participating BSs and the corresponding messages.
\item Eliminate all interfering links except for links from BS-$i_{j}$ to participating UEs in cell $i_{j+1}$,
for all $j \in \langle m \rangle$.
\end{itemize}
We obtain the cyclic (w.r.t cells) IBC modeled as
\begin{equation}
\label{eq:model_converse_cyclic}
Y_{i_{j}}^{[s_{i_{j}}]}(t)  = h_{i_{j}i_{j}}^{[s_{i_{j}}]} X_{i_{j}}(t)
+ h_{i_{j}i_{j-1}}^{[s_{i_{j}}]} X_{i_{j-1}}(t)
+ Z_{i_{j}}^{[s_{i_{j}}]}(t),
\end{equation}
for all  $s_{i_{j}} \in \langle l_{i_{j}} \rangle$ and  $j \in \langle m \rangle$.
As the above steps do not decrease the rates of all remaining messages, we restrict
our attention to the channel in \eqref{eq:model_converse_cyclic} henceforth.
We further define the following side information signal
\begin{equation}
\label{eq:conv_side_inf}
S_{i_{j}}(t)  = h_{i_{j+1}i_{j}}^{[l_{i_{j+1}}]} X_{i_{j}}(t) + Z_{i_{j+1}}^{[l_{i_{j+1}}]}(t), \ j \in \langle m \rangle.
\end{equation}
which is eventually provided to the \emph{stronger} UE in cell $i_{j}$.

For cells $i_{j}$ with $l_{i_{j}} = 1$, Fano's inequality yields
\begin{align}
\label{eq:conv_fano_l1}
n \big(R_{i_{j}}^{[1]} - \epsilon \big)  & \leq I\big(X_{i_{j}}^{n} ; Y_{i_{j}}^{[1]n}, S_{i_{j}}^{n} \big).
\end{align}
On the other hand, for cells $i_{j}$ with $l_{i_{j}} = 2$, we obtain
\begin{align}
\nonumber
n \big( R_{i_{j}}^{[1]} + & R_{i_{j}}^{[2]}  -  2\epsilon \big)   \leq
 I\big(W_{i_{j}}^{[1]} ; Y_{i_{j}}^{[1]n} | W_{i_{j}}^{[2]} \big) \! +  \!  I\big(W_{i_{j}}^{[2]} ; Y_{i_{j}}^{[2]n}  \big) \\
\nonumber
& =  I\big(X_{i_{j}}^{n}, W_{i_{j}}^{[1]} ; Y_{i_{j}}^{[1]n} | W_{i_{j}}^{[2]} \big)
+ I\big(W_{i_{j}}^{[2]} ; Y_{i_{j}}^{[2]n}  \big) \\
\nonumber
& =  I\big(X_{i_{j}}^{n} ; Y_{i_{j}}^{[1]n} | W_{i_{j}}^{[2]} \big)  +
I\big(W_{i_{j}}^{[2]} ; Y_{i_{j}}^{[2]n}  \big) \\
\label{eq:conv_fano_l2_1}
& \leq  I\big(X_{i_{j}}^{n} ; Y_{i_{j}}^{[2]n} | W_{i_{j}}^{[2]} \big) + n +
I\big(W_{i_{j}}^{[2]} ; Y_{i_{j}}^{[2]n}  \big)  \\
\nonumber
& =  I\big(X_{i_{j}}^{n} ; Y_{i_{j}}^{[2]n} \big) + n \\
\label{eq:conv_fano_l2_2}
& \leq  I\big(X_{i_{j}}^{n} ; Y_{i_{j}}^{[2]n}, S_{i_{j}}^{n} \big) + n.
\end{align}
In the above,  \eqref{eq:conv_fano_l2_1} is obtained through a direct application of Lemma \ref{lemma:I_diff},
while taking into consideration the TIN-optimality condition in \eqref{eq:TIN_condition_1}.
By adding the bounds in \eqref{eq:conv_fano_l1} and \eqref{eq:conv_fano_l2_2} for all $i_{j}$, $j \in \langle m \rangle$, we bound the
sum-rate for this cycle as
\begin{multline}
\label{eq:conv_sum_rate_bound}
n \! \! \! \sum_{j \in \langle m \rangle}  \!  \sum_{s_{i_{j}} \in \langle l_{i_{j}} \rangle} \big( R_{i_{j}}^{[s_{i_{j}}]} -  \epsilon \big)
 \! \leq \!  mn \! + \! \sum_{j \in \langle m \rangle}  I\big(X_{i_{j}}^{n} ; Y_{i_{j}}^{[l_{i_{j}}]n}, S_{i_{j}}^{n} \big)  \\
 \leq mn +  n \! \! \! \! \sum_{j \in \langle m \rangle} \! \! \! \log \! \left( \! \! 1 \! +  \! |h_{i_{j}i_{j-1}}^{[l_{i_{j}}]}|^{2}P_{i_{j-1}} \! + \! \frac{|h_{i_{j}i_{j}}^{[l_{i_{j}}]}|^{2}P_{i_{j}}}{1 + |h_{i_{j+1}i_{j}}^{[l_{i_{j+1}}]}|^{2}P_{i_{j}}}  \! \right).
\end{multline}
The bound in \eqref{eq:conv_sum_rate_bound} is obtained by first noting the setting has
essentially reduced to a regular $m$-user IC with receivers given by UE-$(l_{i_{j}},i_{j})$, $j \in \langle m \rangle$,
and side information signals as defined in \eqref{eq:conv_side_inf}, and then applying the steps in \cite[Appendix C]{Geng2015}.
Combining \eqref{eq:conv_sum_rate_bound} with the condition in \eqref{eq:TIN_condition_2},
the corresponding GDoF inequality in \eqref{eq:IBC_TIN_region_3} is obtained (after rearranging indices).
\begin{remark}
The capacity outer bound in \eqref{eq:conv_sum_rate_bound} leads to a constant-gap characterization of the capacity
region $\mathcal{C}^{\mathrm{IBC}}$ when the TIN-optimality conditions in Theorem \ref{theorem:TIN_optimality} hold.
\end{remark}
\section{Conclusion}
In this paper, we established an equivalence between the
achievable GDoF regions for the IBC and IMAC
under power controlled single-cell transmissions, while treating inter-cell
interference as noise.
This uplink-downlink duality, on the achievability side,
leads to a characterization of the IBC TIN-achievable GDoF region.
We also identified a regime in which the IBC TIN-achievable GDoF region is optimal.
This IBC TIN-optimal regime is included in its counterpart IMAC TIN-optimal regime in \cite{Joudeh2018a}.
It is of interest to investigate whether the IBC TIN-optimal regime in Theorem \ref{theorem:TIN_optimality} can be enlarged to
coincide with the IMAC TIN-optimal regime in \cite{Joudeh2018a}.
\appendices
\section{Proof of Theorem \ref{theorem:TIN_UL_DL_Duality}}
\label{appebdix:proof_of_duality}
\subsection{$ \mathcal{P}^{\mathrm{IBC}\star} \subseteq  \mathcal{P}^{\mathrm{IMAC}\star} $}
Consider $\mathbf{d} \in \mathcal{P}_{\bm{\pi}}^{\mathrm{IBC}\star}(\mathbf{r})$ with arbitrary $(\bm{\pi},\mathbf{r})$.
We observe that \eqref{eq:IBC_GDoF per user} is equivalently expressed as
\begin{equation}
\label{eq:duality_IBC_GDoF_TIN}
0 \! \leq \! d_{k}^{[\pi_{k}(l_{k})]} \! \leq \!  \max \! \left\{ \! 0, r_{k}^{[\pi_{k}(l_{k})]} + \beta^{[\pi_{k}(l_{k})]} \!
 \right\}, l_{k} \in \{1,2\}
\end{equation}
where $\beta^{[\pi_{k}(l_{k})]}$, $l_{k} \in \{1,2\}$, is given by
\begin{subequations}
\begin{align}
\nonumber
\beta_{k}^{[\pi_{k}(1)]}   = &  \min_{m_{k} \in \{1,2\}} \min  \Bigl\{ \alpha_{kk}^{[\pi_{k}(m_{k})]},
 -  r_{k}^{[\pi_{k}(2)]}   ,   \\
 & \alpha_{kk}^{[\pi_{k}(m_{k})]} \! -   \!
 \max_{(l_{j},j) : j \neq k} \big\{ ( \alpha_{kj}^{[\pi_{k}(m_{k})]}  +
 r_{j}^{[l_{j}]} ) \big\}  \Bigr\}  \\
\nonumber
\beta _{k}^{[\pi_{k}(2)]}  = &  \min \Bigl\{ \alpha_{kk}^{[\pi_{k}(2)]} , \\
& \alpha_{kk}^{[\pi_{k}(2)]} - \max_{ (l_{j},j): j\neq k} \big\{( \alpha_{kj}^{[\pi_{k}(2)]} + r_{j}^{[l_{j}]} ) \big\} \Bigr\}.
\end{align}
\end{subequations}
Now consider the following power allocation for the IMAC
\begin{equation}
\label{eq:r_duality_IMAC}
\bar{r}_{k}^{[\pi_{k}(l_{k})]}   = - \alpha_{kk}^{[\pi_{k}(l_{k})]} + \beta_{k}^{[\pi_{k}(l_{k})]}.
\end{equation}
As $\beta_{k}^{[\pi_{k}(1)]} \leq \min\big\{ \alpha_{kk}^{[\pi_{k}(1)]}, \alpha_{kk}^{[\pi_{k}(2)]} \big\}$
and $\beta_{k}^{[\pi_{k}(2)]} \leq \alpha_{kk}^{[\pi_{k}(2)]} $, the power allocation in
\eqref{eq:r_duality_IMAC} is feasible.
Using \eqref{eq:r_duality_IMAC}, we achieve the set of IMAC GDoF tuples
$\mathcal{P}_{\bm{\pi}}^{\mathrm{IMAC}\star}(\bar{\mathbf{r}})$ that satisfy
\begin{multline}
\label{eq:duality_IMAC_GDoF_TIN_IBC}
\bar{d}_{k}^{[\pi_{k}(l_{k})]}  \! \leq \!
\max \biggl\{ \!  0,   \beta_{k}^{[\pi_{k}(l_{k})]}
 \!  -  \! \Bigl(\max \Bigl\{ \beta_{k}^{[\pi_{k}(1)]} \mathbbm{1}(l_{k} = 2), \\
  \max_{(l_{j},j): j \neq k} \{ \alpha_{jk}^{[l_{j}]} + \bar{r}_{j}^{[l_{j}]}   \}   \Bigr\} \Bigr)^{+}   \biggr\}.
\end{multline}
From \eqref{eq:duality_IBC_GDoF_TIN} and  \eqref{eq:duality_IMAC_GDoF_TIN_IBC},
to show that  $\mathbf{d}$ is also in $\mathcal{P}_{\bm{\pi}}^{\mathrm{IMAC}\star}(\bar{\mathbf{r}})$, it is sufficient to show that
\begin{equation}
\label{eq:duality_IMAC_GDoF_UB_1}
r_{k}^{[\pi_{k}(l_{k})]} \!  \! \leq   \! \min \Bigl\{ \!  0,  \! - \beta_{k}^{[\pi_{k}(1)]}  \mathbbm{1}( l_{k}  \!  \! =  \! 2  ),
 \!  \!  \min_{(l_{j},j): j \neq k}  \!  \!  \! \big\{\! -(\alpha_{jk}^{[l_{j}]} + \bar{r}_{j}^{[l_{j}]})  \! \big\} \! \Bigr\}.
\end{equation}
Since $r_{k}^{[\pi_{k}(l_{k})]}  \leq 0$, we only need to show that the inequality in \eqref{eq:duality_IMAC_GDoF_UB_1}
holds for the two other terms in the $\min\{0,\cdot,\cdot\}$.
We start with $-\beta_{k}^{[\pi_{k}(1)]}$, which only has an influence when $l_{k} = 2$,
\begin{align}
\nonumber
- \beta_{k}^{[\pi_{k}(1)]}  = & \max_{m_{k} \in \{1,2\}} \max  \Bigl\{ -\alpha_{kk}^{[\pi_{k}(m_{k})]},
   r_{k}^{[\pi_{k}(2)]}   , \\
\nonumber
& - \alpha_{kk}^{[\pi_{k}(m_{k})]} + \max_{(l_{j},j) : j \neq k} \big\{  \alpha_{kj}^{[\pi_{k}(m_{k})]}  +   r_{j}^{[l_{j}]}  \big\}  \Bigr\} \\
   \label{eq:duality_IMAC_GDoF_UB_2}
    \geq &  r_{k}^{[\pi_{k}(2)]}.
\end{align}
Next, we observe that for any $(l_{j},j)$ with $j \neq k$, we have
\begin{align}
\nonumber
- ( \alpha_{jk}^{[l_{j}]} + \bar{r}_{j}^{[l_{j}]}  ) & = - \alpha_{jk}^{[l_{j}]} + \alpha_{jj}^{[l_{j}]} - \beta_{j}^{[l_{j}]}  \\
\nonumber
&\geq  -\alpha_{jk}^{[l_{j}]} +  \max_{(l_{i},i):i \neq j} \{ \alpha_{ji}^{[l_{j}]} + r_{i}^{[l_{i}]}   \}  \\
\label{eq:duality_IMAC_GDoF_UB_5}
& \geq r_{k}^{[\pi_{k}(l_{k})]}
\end{align}
From \eqref{eq:duality_IMAC_GDoF_UB_2} and \eqref{eq:duality_IMAC_GDoF_UB_5}, we conclude that the inequality in \eqref{eq:duality_IMAC_GDoF_UB_1} holds. This in turn implies that $\mathcal{P}_{\bm{\pi}}^{\mathrm{IBC}\star}(\mathbf{r}) \subseteq \mathcal{P}_{\bm{\pi}}^{\mathrm{IMAC}\star}(\bar{\mathbf{r}})$, and hence completes this part of the proof.
\subsection{$ \mathcal{P}^{\mathrm{IMAC}\star} \subseteq  \mathcal{P}^{\mathrm{IBC}\star} $}
Consider $\bar{\mathbf{d}} \in \mathcal{P}_{\bm{\pi}}^{\mathrm{IMAC}\star}(\bar{\mathbf{r}})$ with arbitrary $(\bm{\pi},\bar{\mathbf{r}})$.
We start by observing that, without loss of generality, we may assume
\begin{equation}
\label{eq:duality_condition_IMAC_order}
\bar{r}_{k}^{[\pi_{k}(2)]} +  \alpha_{kk}^{[\pi_{k}(2)]} \geq  \bar{r}_{k}^{[\pi_{k}(1)]} +  \alpha_{kk}^{[\pi_{k}(1)]}, \
k \in \langle K \rangle,
\end{equation}
as otherwise, there exists another satisfactory power allocation $\tilde{\mathbf{r}}$ 
with an achievable GDoF tuple such that $\tilde{\mathbf{d}} \geq \bar{\mathbf{d}}$ 
(see end of this section).
Next, we observe that \eqref{eq:IMAC_GDoF per user} can be expressed by
\begin{equation}
\label{eq:duality_IMAC_GDoF_beta}
0 \leq \bar{d}_{k}^{[\pi_{k}(l_{k})]}  \leq
\max \biggl\{  0,   \alpha_{kk}^{[\pi_{k}(l_{k})]} +  \bar{r}_{k}^{[\pi_{k}(l_{k})]}
  -  \gamma_{k}^{[\pi_{k}(l_{k})]}  \biggr\}
\end{equation}
where we define $\gamma_{k}^{[\pi_{k}(l_{k})]}$, $l_{k} \in \{1,2\}$, as follows
\begin{multline}
\label{eq:duality_IMAC_GDoF_beta_def}
\gamma_{k}^{[\pi_{k}(l_{k})]} =
\max \Bigl\{ 0, (\alpha_{kk}^{[\pi_{k}(1)]} + \bar{r}_{k}^{[\pi_{k}(1)]}) \mathbbm{1}(l_{k} = 2) , \\
  \max_{(l_{j},j): j \neq k} \{ \alpha_{jk}^{[l_{j}]} + \bar{r}_{j}^{[l_{j}]}   \}   \Bigr\}.
\end{multline}
Now consider the IBC power allocation given by
\begin{equation}
\label{eq:r_duality_IBC}
r_{k}^{[\pi_{k}(l_{k})]} =  - \gamma_{k}^{[\pi_{k}(l_{k})]}
\end{equation}
which is clearly feasible.
Using \eqref{eq:r_duality_IBC}, we achieve the set of IBC GDoF tuples
$\mathcal{P}_{\bm{\pi}}^{\mathrm{IBC}\star}(\bar{\mathbf{r}})$ that satisfy
\begin{subequations}
\label{eq:duality_IBC_GDoF_beta}
\begin{align}
\nonumber
d_{k}^{[\pi_{k}(1)]}  \leq  \max & \biggl\{  0,   \min_{m_{k} \in \{1,2\} }    \Bigl\{ \alpha_{kk}^{[\pi_{k}(m_{k})]}   -  \gamma_{k}^{[\pi_{k}(1)]}
 - \\
\Bigl( \! \max \bigl\{ \! \alpha_{kk}^{[\pi_{k}(m_{k})]} & \! - \! \gamma_{k}^{[\pi_{k}(2)]} ,
\max_{(l_{j},j):j \neq k} \{ \alpha_{kj}^{[\pi_{k}(m_{k})]}
\! - \! \gamma_{j}^{[l_{j}]} \} \! \bigr\} \! \Bigr)^{+} \! \Bigr\} \!  \biggr\} \\
\nonumber
d_{k}^{[\pi_{k}(2)]}  \leq   \max & \biggl\{ 0, \alpha_{kk}^{[\pi_{k}(2)]} - \gamma_{k}^{[\pi_{k}(2)]} \\
&  -  \Bigl( \max_{(l_{j},j): j \neq k} \{   \alpha_{kj}^{[\pi_{k}(2)]} - \gamma_{j}^{[l_{j}]}  \} \Bigr)^{+} \biggr\}.
\end{align}
\end{subequations}
By examining  \eqref{eq:duality_IMAC_GDoF_beta} and \eqref{eq:duality_IBC_GDoF_beta}, $\mathcal{P}_{\bm{\pi}}^{\mathrm{IMAC}\star}(\mathbf{r}) \subseteq \mathcal{P}_{\bm{\pi}}^{\mathrm{IBC}\star}(\bar{\mathbf{r}})$
is shown by proving that the following inequalities hold
\begin{subequations}
\label{eq:duality_IBC_GDoF_UB_2}
\begin{align}
\nonumber
\alpha_{kk}^{[\pi_{k}(1)]} + \bar{r}_{k}^{[\pi_{k}(1)]}  & \leq  \min_{m_{k} \in \{1,2\} }  \min  \Bigl\{
\alpha_{kk}^{[\pi_{k}(m_{k})]} ,  \gamma_{k}^{[\pi_{k}(2)]} , \\
\label{eq:duality_IBC_GDoF_UB_2_a}
 \alpha_{kk}^{[\pi_{k}(m_{k})]} + & \min_{(l_{j},j):j \neq k} \{\gamma_{j}^{[l_{j}]} -
  \alpha_{kj}^{[\pi_{k}(m_{k})]} \}    \Bigr\} \\
\nonumber
\alpha_{kk}^{[\pi_{k}(2)]} + \bar{r}_{k}^{[\pi_{k}(2)]} & \leq   \min \Bigl\{ \alpha_{kk}^{[\pi_{k}(2)]} , \\
\label{eq:duality_IBC_GDoF_UB_2_b}
\alpha_{kk}^{[\pi_{k}(2)]} +  &  \min_{(l_{j},j): j \neq k}  \{ \gamma_{j}^{[l_{j}]} - \alpha_{kj}^{[\pi_{k}(2)]} \}  \Bigr\}.
\end{align}
\end{subequations}
We start by showing that \eqref{eq:duality_IBC_GDoF_UB_2_a} holds.
For the first of the three terms inside the $\min_{m_{k}}\min \{\cdot,\cdot,\cdot\}$ in \eqref{eq:duality_IBC_GDoF_UB_2_a},
we note that $\alpha_{kk}^{[\pi_{k}(1)]} + \bar{r}_{k}^{[\pi_{k}(1)]} \leq \alpha_{kk}^{[\pi_{k}(m_{k})]}$, $m_{k} \in \{1,2\}$,
holds due to \eqref{eq:duality_condition_IMAC_order}.
For the second term, we observe that $\alpha_{kk}^{[\pi_{k}(1)]} + \bar{r}_{k}^{[\pi_{k}(1)]} \leq \gamma_{k}^{[\pi_{k}(2)]}$
follows directly from \eqref{eq:duality_IMAC_GDoF_beta_def}.
For the final term, we observe that for any $(l_{j},j) \in \mathcal{K}$, such that $ j \neq k$, we have
\begin{align}
\nonumber
\alpha_{kk}^{[\pi_{k}(m_{k})]} + (\gamma_{j}^{[l_{j}]}  -  \alpha_{kj}^{[\pi_{k}(m_{k})]})   &  \geq
\alpha_{kk}^{[\pi_{k}(m_{k})]} +  \\
\label{eq:duality_IBC_GDoF_UB_5}
 \max_{(l_{i},i): i \neq j} \{\alpha_{ij}^{[l_{i}]} & + \bar{r}_{i}^{[l_{i}]} \} - \alpha_{kj}^{[\pi_{k}(m_{k})]}  \\
\label{eq:duality_IBC_GDoF_UB_6}
 \geq & \ \alpha_{kk}^{[\pi_{k}(m_{k})]}  + \bar{r}_{k}^{[\pi_{k}(m_{k})]}   \\
\label{eq:duality_IBC_GDoF_UB_7}
\geq & \ \alpha_{kk}^{[\pi_{k}(1)]}  + \bar{r}_{k}^{[\pi_{k}(1)]}
\end{align}
where \eqref{eq:duality_IBC_GDoF_UB_6} is obtained by setting
$(l_{i},i) = (\pi_{k}(m_{k}),k)$ in  \eqref{eq:duality_IBC_GDoF_UB_5},
and the inequality in \eqref{eq:duality_IBC_GDoF_UB_7} holds due to \eqref{eq:duality_condition_IMAC_order}.
Next, we consider \eqref{eq:duality_IBC_GDoF_UB_2_b}. It is clear that the inequality holds for the first of the two
terms in the $\min\{\cdot, \cdot\}$.
The inequality also holds for the second term in the $\min\{\cdot, \cdot\}$ due to \eqref{eq:duality_IBC_GDoF_UB_6}.
As \eqref{eq:duality_IBC_GDoF_UB_2} holds, we have $\mathcal{P}_{\bm{\pi}}^{\mathrm{IMAC}\star}(\mathbf{r}) \subseteq \mathcal{P}_{\bm{\pi}}^{\mathrm{IBC}\star}(\bar{\mathbf{r}})$, which concludes the proof.
\subsection{Justification for \eqref{eq:duality_condition_IMAC_order}}
To show that the assumption in \eqref{eq:duality_condition_IMAC_order} is justified, suppose that the contrary, i.e. $\bar{r}_{k}^{[\pi_{k}(2)]} +  \alpha_{kk}^{[\pi_{k}(2)]} <  \bar{r}_{k}^{[\pi_{k}(1)]} +  \alpha_{kk}^{[\pi_{k}(1)]}$, holds for some $k \in \langle K \rangle$.
From \eqref{eq:IMAC_GDoF per user_2}, it follows that in this case we would necessarily have $\bar{d}_{k}^{[\pi_{k}(2)]} = 0$.
Now consider an alternative scheme $(\tilde{\bm{\pi}},\tilde{\mathbf{r}})$, which is a modification of $(\bm{\pi},\bar{\mathbf{r}})$ such that: $\tilde{\pi}_{k}(1) = \pi_{k}(2)$ and $\tilde{\pi}_{k}(2) = \pi_{k}(1)$ (i.e. swapping the order in cell $k$), and $\bar{r}_{k}^{[\tilde{\pi}_{k}(1)]} = -\infty$,
while maintaining the order and power allocation for all remaining cells.
For cell $k$, this modified scheme achieves any GDoF pair satisfying
$\tilde{d}_{k}^{[\tilde{\pi}_{k}(1)]}  = 0$ and
$0 \leq \tilde{d}_{k}^{[\tilde{\pi}_{k}(2)]}   \leq
\max \bigl\{  0,   \alpha_{kk}^{[\tilde{\pi}_{k}(2)]}  +   r_{k}^{[\tilde{\pi}_{k}(2)]}
 -  (\max_{(l_{j},j): j \neq k} \{ \alpha_{jk}^{[l_{j}]} + r_{j}^{[l_{j}]}   \}    )^{+}   \bigr\}$.
Recalling that the decoding order is swapping in cell $k$,
it follows that all GDoF pairs of cell $k$ achieved using $(\bm{\pi},\bar{\mathbf{r}})$
are also achievable using $(\tilde{\bm{\pi}},\tilde{\mathbf{r}})$.
Moreover, all GDoF pairs of all remaining cells
achieved using $(\bm{\pi},\bar{\mathbf{r}})$ are also achievable using $(\tilde{\bm{\pi}},\tilde{\mathbf{r}})$.
This last statement holds as the power allocation for cells $i \in \langle K \rangle \setminus \{k\}$ is unaltered, while the transmit power of cell $k$ is reduced in $(\tilde{\bm{\pi}},\tilde{\mathbf{r}})$.
Therefore, we have  $\mathcal{P}_{\bm{\pi}}^{\mathrm{IMAC}\star}(\bar{\mathbf{r}}) \subseteq \mathcal{P}_{\tilde{\bm{\pi}}}^{\mathrm{IMAC}\star}(\tilde{\mathbf{r}})$.
We apply the argument to all cells $k$ that violate \eqref{eq:duality_condition_IMAC_order}.
\section{Proof of Lemma \ref{lemma:I_diff}}
\label{appebdix:proof_lemma_I_diff}
First, we observe that $I(X_{1}^{n} ; Y_{a}^{n} ) $ is bounded above as
\begin{align}
\label{eq:I_diff_proof_UB}
I(X_{1}^{n} ; Y_{a}^{n} ) \leq I(X_{1}^{n} ; a_{1}X_{1}^{n} + Z_{a}^{n} )
\end{align}
which follows from the independence of $X_{1}^{n}$ and $X_{2}^{n}$.
Next, we bound $I(X_{1}^{n} ; Y_{b}^{n} )$ below as
\begin{align}
I(X_{1}^{n} ; Y_{b}^{n} ) & = I\Big(X_{1}^{n} ; \frac{b_{1}}{b_{2}}X_{1}^{n} + X_{2}^{n} + \frac{1}{b_{2}}Z_{b}^{n} \Big) \\
\label{eq:I_diff_proof_LB_1}
& \geq I\Big(X_{1}^{n} ; \frac{b_{1}}{b_{2}}X_{1}^{n} + X_{2}^{n} + Z_{b}^{n} \Big) \\
\label{eq:I_diff_proof_LB_2}
& \geq I\Big( X_{1}^{n} ; \frac{b_{1}}{b_{2}}X_{1}^{n} \! + \! Z_{b}^{n}  \Big) \! - \! I( X_{2}^{n} ; X_{2}^{n} \! + \!  Z_{b}^{n}
) \\
\label{eq:I_diff_proof_LB_3}
& \geq I (X_{1}^{n} ; a_{1}X_{1}^{n} \! + \! Z_{b}^{n} ) \!-\! I(X_{2}^{n} ; X_{2}^{n} \! + \! Z_{b}^{n} ).
\end{align}
Inequality \eqref{eq:I_diff_proof_LB_1} holds due to $|b_{2}|^{2} \geq 1$ and $Z_{b}(t) \sim\mathcal{N}_{\mathbb{C}}(0,1)$, while
\eqref{eq:I_diff_proof_LB_2} is obtained from  the chain rule as follows
\begin{align}
\nonumber
&I\Big(X_{1}^{n},X_{2}^{n} ; \frac{b_{1}}{b_{2}}X_{1}^{n} + X_{2}^{n} + Z_{b}^{n} \Big)\\
&= I\Big(X_{1}^{n} ; \frac{b_{1}}{b_{2}}X_{1}^{n} + X_{2}^{n} + Z_{b}^{n} \Big)  +
I\Big(X_{2}^{n} ; X_{2}^{n} + Z_{b}^{n} \Big)   \\
&= I\Big(X_{1}^{n} ; \frac{b_{1}}{b_{2}}X_{1}^{n} \! + \! Z_{b}^{n} \Big)   +
I\Big(X_{2}^{n} ; \frac{b_{1}}{b_{2}}X_{1}^{n} \! +  \! X_{2}^{n} \! +  \! Z_{b}^{n} \Big)  \\
& \geq I\Big(X_{1}^{n} ; \frac{b_{1}}{b_{2}}X_{1}^{n} + Z_{b}^{n} \Big).
\end{align}
On the other and, inequality \eqref{eq:I_diff_proof_LB_3} holds since $\frac{ |b_{1}|^{2} }{ |b_{2}|^{2}  } \geq |a_{1}|^{2}$.
From \eqref{eq:I_diff_proof_UB} and \eqref{eq:I_diff_proof_LB_3}, we obtain
\begin{equation}
\label{eq:lemma_I_diff_no_W}
I(X_{1}^{n} ; Y_{a}^{n} ) - I(X_{1}^{n} ; Y_{b}^{n} )  \leq I(X_{2}^{n} ; X_{2}^{n} + Z_{b}^{n} ) \overset{\text{(a)}}{\leq} n
\end{equation}
where (a) in \eqref{eq:lemma_I_diff_no_W} is obtained by taking $X_{2}^{n} \sim \mathcal{N}_{\mathbb{C}}(\mathbf{0},\mathbf{I}_{n})$.
Equipped with \eqref{eq:lemma_I_diff_no_W}, we obtain \eqref{eq:lemma_I_diff}
as follows
\begin{align}
\nonumber
& I(X_{1}^{n} ; Y_{a}^{n} |W) -  I(X_{1}^{n} ; Y_{b}^{n}|W)  \\
\label{eq:lemma_I_diff_W_1}
& =  \sum_{w} p(w)
\Big[  I\big(X_{1}^{n} ; Y_{a}^{n} |W \! = \! w \big) \! - \! I\big(X_{1}^{n} ; Y_{b}^{n}|W \! = \! w \big)  \Big]\\
\label{eq:lemma_I_diff_W_2}
& = \sum_{w}p(w)
\Big[ I\big(X_{1w}^{n} ; Y_{a}^{n}  \big) - I\big(X_{1w}^{n} ; Y_{b}^{n} \big) \Big] \overset{\text{(b)}}{\leq} n
\end{align}
where $X_{1w}^{n} \sim X_{1}^{n}|\{W = w\} \sim F(x_{1}^{n} | w)$.
The transition from \eqref{eq:lemma_I_diff_W_1} to \eqref{eq:lemma_I_diff_W_2}  is due to the Markov chains
$W \rightarrow X_{1}^{n} \rightarrow Y_{a}^{n}$ and $W \rightarrow X_{1}^{n} \rightarrow Y_{b}^{n}$ (see \cite[Ch. 5.6.1]{ElGamal2011}).
On the other hand, (b) in \eqref{eq:lemma_I_diff_W_2} follows from  \eqref{eq:lemma_I_diff_no_W}, which completes the proof.
\bibliographystyle{IEEEtran}
\bibliography{References}
\end{document}